\documentclass{article}

\usepackage{arxiv}
\usepackage{graphicx}
\usepackage{amssymb}
\usepackage[utf8]{inputenc} 
\usepackage[T1]{fontenc}    
\usepackage{hyperref}       
\usepackage{url}            
\usepackage{booktabs}       
\usepackage{amsfonts}       
\usepackage{nicefrac}       
\usepackage{microtype}      
\usepackage{lipsum}

\title{Suppressed self-discharge of an aqueous supercapacitor using Earth-abundant materials}

\author{
  Samuel Devese\\
  School of Chemical and Physical Sciences\\
  Victoria University of Wellington\\
  Kelburn, 6140, New Zealand \\
  \texttt{sam.devese@vuw.ac.nz} \\
   \And
 Thomas Nann \\
  School of Mathematical and Physical Sciences\\
  The University of Newcastle\\
  Callaghan, 2308 NSW, Australia\\
  \texttt{thomas.nann@newcastle.edu.au} \\
}

\begin{document}
\maketitle

\begin{abstract}
Supercapacitors (aka electrostatic double-layer capacitors --- EDLCs) offer excellent power storage capacity and kinetics, but suffer under rapid self-discharge. We introduced a zeolite framework into the active capacitor electrode, with the goal to tailor the free desorption energy and thus the self-discharge rate of supercapacitors. Low-cost carbonaceous materials and benchtop production methods were used to create supercapacitor electrodes with a measured specific capacitance of 17.25 F g$^{-1}$, a coulombic efficiency of 100\%, and charge retention of over 25\% over 24 hours determined by galvanostatic charge/discharge curve measurements. This charge retention was an enhancement of $\sim$350\% compared with electrodes without zeolite coating.
\end{abstract}


\section{Introduction}
\label{S:1}
With the increasing electrification of appliances throughout all aspects of our lives, we are becoming more and more dependent on portable energy storage. There is a diversity of applications that all require different energy storage specifications. At present, this is mainly achieved via lithium ion batteries \cite{blomgren_development_2017, nitta_li-ion_2015} owing to their high energy density and well-established battery chemistry. Conversely, supercapacitors have a high power density, able to supply large amounts of power for short periods of time \cite{halper_supercapacitors_2006, wang_engineering_2015, zhang_progress_2009}.

While batteries store energy via chemical reactions, supercapacitors store energy by forming a layer of electrolytic ions on the surface of each electrode (electrostatic double-layer capacitor --- EDLC). This physical adsorption of ions results in very little degradation of electrodes, and so supercapacitors have excellent cyclability in the order of millions of cycles \cite{bittner_ageing_2012, lewandowski_self-discharge_2013, cohn_durable_2016, zhang_review_2011}. It is worth noting that, due to the mechanism by which they store energy, supercapacitors have an inferior energy density compared with batteries. For applications with extended runtimes, batteries are the preferred option, but it has been shown that supercapacitors can complement energy storage systems through hybridisation, as well as excelling in specific high-power, short-duration applications \cite{herrera_adaptive_2016, wei_discussion_2009, smith_using_2002, pay_effectiveness_2003, stepanov_development_nodate}.

State of the art supercapacitor electrodes such as RuO\textsubscript{2} and MnO\textsubscript{2} have theoretical maximum capacitance values of 1360 and 1370 F g\textsuperscript{-1} respectively \cite{wang_review_2012, toupin_charge_2004}. However, supercapacitors have more fundamental issues that need addressing to increase their use as energy storage devices; namely their high production cost and their tendency to self-discharge \cite{halper_supercapacitors_2006, andreas_self-discharge_2015, conway_diagnostic_1997, ricketts_self-discharge_2000}. Electrode materials such as MnO\textsubscript{2} and other transition metal oxides increase the fabrication cost and performance can depend heavily on their crystal structure \cite{long_voltammetric_1999}. Carbon materials feature much more stability \cite{yang_-situ_2019} \cite{frackowiak_carbon_2001} and can perform with competitive specific capacitance values depending on the fabrication method \cite{fan_high_2007, haladkar_preparation_2018, lu_electrospun_2017}. In the context of this study, carbon slurries are worth noting, exhibiting capacitances between 0.5 and 85 F g\textsuperscript{-1} \cite{akgul_characterization_2016, kadir_application_2011, abbas_strategies_2015, chmiola_anomalous_2006}.

There are many factors that determine the suitability of an electrolyte for the use with supercapacitor electrodes. Zhong \textit{et al.} identified nine properties that an electrolyte would ideally possess, summarised here in Table \ref{electrolytetable} \cite{zhong_review_2015, beguin_carbons_2014}. For example, ionic liquids and organic electrolytes are costly but exhibit a wide voltage window of up to 7 V, allowing for a subsequently higher working capacitance \cite{hayyan_investigating_2013, ignatev_new_2005}. Aqueous electrolytes can be seen to exhibit the most ideal properties, suffering only under a relatively small voltage window of about 1.2 V. However, it has been reported that over 80\% of supercapacitor literature used an aqueous electrolyte \cite{zhong_review_2015, tanahashi_comparison_1990}.

\begin{table}[h]
\centering
\begin{tabular}{l l l l}
\hline
\textbf{Property} & \textbf{Ionic} & \textbf{Organic} & \textbf{Aqueous}\\
\hline
 Wide voltage window & \checkmark & \checkmark & $\times$ \\
 Wide temperature window & $\times$ & $\times$ & \checkmark \\
 High ionic conductivity & $\times$ & $\times$ & \checkmark \\
 High stability & \checkmark & \checkmark & \checkmark \\
 Well matched to (carbon) electrodes & $\times$ & \checkmark & \checkmark \\
 Does not react to electrodes & $\times$ & \checkmark & \checkmark \\
 Low volatility & \checkmark & $\times$ & \checkmark \\
 Low cost & $\times$ & $\times$ & \checkmark \\
 Environmentally friendly & \checkmark & $\times$ & \checkmark \\
\hline
\end{tabular}\caption{Comparison of how each electrolyte type displays ‘ideal’ behaviour, based on nine important properties for an ideal electrolyte, as stated by Zhong \textit{et al.} \cite{zhong_review_2015}.}
\label{electrolytetable}
\end{table}

\begin{figure}[hbt]
\centering\includegraphics[width=0.85\linewidth]{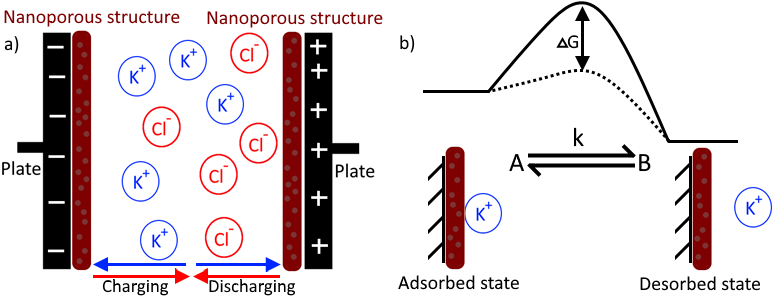}
\caption{a) Schematic of the constructed supercapacitor, shown with carbon paper as the plate, zeolite slurry as the nanoporous structure and potassium chloride as the electrolyte. b) Energy profile for the desorption of a potassium ion with (solid line) and without (dotted line) a zeolite slurry coating.}
\label{fig_scheme}
\end{figure}

In this work, the fundamental issues with supercapacitors were addressed by designing a conceptually different cell architecture. Charging a supercapacitor causes ions to adsorb onto the electrode surfaces as schematically depicted in Figure \ref{fig_scheme}a. When no more ions are able to be adsorbed, the device is fully charged. In order to desorb from the electrodes, the ions have to overcome a free energy barrier as displayed in Figure \ref{fig_scheme}b (dotted line). This barrier has to be low enough, so that ions can desorb when needed (\textit{viz.}  during discharge), but high enough to prevent spontaneous self-discharge (leaking) \cite{andreas_self-discharge_2015, conway_diagnostic_1997, ricketts_self-discharge_2000}. We hypothesise that the height of this critical barrier can be increased by introducing nanoporous materials such as zeolites. We propose that these nanopores provide an additional energy barrier, thus lowering the rate of ion diffusion and, subsequently, self-discharge. The Linde Type A (LTA) framework can accommodate particles up to a size of 4.21 Å and is readily available in the form of molecular sieves \cite{noauthor_database_nodate}. A potassium chloride electrolyte was chosen due to the favourable characteristics of aqueous electrolytes in general and specifically due to the similarity in hydrated ionic diameters of each ion (K\textsuperscript{+}: 3.31 Å and Cl\textsuperscript{-}: 3.32 Å) \cite{zhong_review_2015}, as well as compatibility with the zeolite. We were able to demonstrate that the introduction of this physical barrier reduced the self-discharge time and allows for future construction of new types of supercapacitors using very simple, inexpensive materials.

\section{Materials and methods}
\label{S:2}

Carbon black (Super P), polyvinylidene fluoride (PVDF) (600,000 g mol\textsuperscript{-1}), N-methyl-2-pyrrolidone (NMP), 4 Å molecular sieves, potassium chloride powder, molybdenum metal sheet, and Whatman\textsuperscript{\textregistered} Glass Microfiber Filters (Grade GF/F) were used as obtained from Sigma-Aldrich.

\subsection{Electrode preparation}

A 70 wt\% zeolite slurry solution was prepared by adding NMP dropwise until it completely dissolved 0.05 g of PVDF (about 2 mL). To this solution, 0.10 g of carbon black and 0.35 g of 4 Å molecular sieves were added, which formed a viscous black slurry that was left magnetically stirring overnight.

The slurry was applied to a conductive carbon paper surface using the doctor blade method. This is also known as tape-casting, as the surface is taped down with an exposed central area. A 3.5x3.5 cm square was taped down with a 1.5x1.5 cm exposed area and the slurry was dropped along the one side of the exposed area and then smoothed over the entire surface using a microscope slide. This was left to dry at room temperature for 72 hours, until the NMP had evaporated and left a dry, black coating that was approximately as thick as the tape (about 0.2 mm). Some cracking was observed but, in general, the coating was continuous, and the electrodes used were taken from areas with no cracking.

\subsection{Supercapacitor assembly}

The process of constructing a supercapacitor cell from the slurry coated carbon paper electrodes involved punching out two electrode discs with a diameter of 1 cm with individual slurry-loadings of 0.75 and 0.91 mg, for a total electrode mass of 1.66 mg. Supercapacitor cells were constructed using a custom-made polyether ether ketone (PEEK) body casing with separator, electrode, and current collector discs having a diameter of 1 cm (refer to supporting information). 7--8 drops of 3M KCl electrolyte were pipetted onto two glass microfibre separator discs (Whatman\textsuperscript{\textregistered} Glass Microfiber Filters); just enough to fully saturate them. The current collector discs were punched from 0.1 mm thick molybdenum sheets, and the entire cell was constructed on the benchtop in open air. Supercapacitors were similarly assembled, with 1 cm discs of uncoated carbon paper (total electrode mass of 2.80 mg) as electrodes for comparison.

\subsection{Supercapacitor characterisation}

The two-electrode cell setup contained in the PEEK body casing provided a more realistic and accurate way of determining performance values for supercapacitor electrodes than a three-electrode setup, as it replicated the conditions of a working supercapacitor, noted in the best practice paper by Stoller and Ruoff \cite{d.stoller_best_2010}. They go on to state that galvanostatic charge-discharge curves best reflect how a cell would perform in practical use if a constant current was used for the entire discharge, allowing for accurate capacitance calculations.

Capacitance can be calculated from the discharge curve using Equation \ref{eq:Cap1}, where $C$ is cell capacitance (F), $I$ is the applied current (A), and $\frac{dV}{dt}$ is the slope of the discharge curve (V s\textsuperscript{-1}).

\begin{equation}
C=\frac{I}{\frac{dV}{dt}}
\label{eq:Cap1}
\end{equation}

This is the capacitance of the whole cell which can be converted to specific capacitance of a single electrode using Equation \ref{eq:Cap2}, with $C_{sp}$ as specific capacitance (F g$^{-1}$), and $m$ as mass (g) of the total active material for both electrodes. The multiplier of 4 is a result of two doubling effects; firstly to account for the halved voltage due to use of a two-electrode cell and secondly to separate the entire cell capacitance into individual electrode capacitances. \cite{d.stoller_best_2010, vijayakumar_electrode_2019, zhu_carbon-based_2011, taberna_electrode_2006}

\begin{equation}
C_{sp}=\frac{4C}{m}
\label{eq:Cap2}
\end{equation}

Using the constructed supercapacitor cells, data for charge-discharge curves were obtained by connecting the PEEK cells to a Neware BTS4000 battery analyser and controlling both voltage and current from their software. Prior to data collection, a pre-measurement cycle was performed to flush the electrodes with electrolyte and ions that involved a short, 10-cycle charge/discharge process between 0.1--1.1 V \cite{lehtimaki_performance_2017}. For measurements, cells were typically cycled between 0.2 and 1.2 V at a constant current of 0.5 A g\textsuperscript{-1} based on active material mass. Other voltage ranges and charging currents were also investigated. A charge rate of 0.5 A g\textsuperscript{-1} gave a charge/discharge time that would be realistic for actual supercapacitor use with the slurry coated electrodes, but this was found to be too high of a charge rate for the uncoated carbon paper electrodes. Instead, they were cycled with a current of 0.1 A g\textsuperscript{-1}.

Lifetime and stability are important parameters for supercapacitor electrodes, and so cells were cycled for a number of different charge/discharge cycles (CDCs), including 10, 1000, 5000, and 10000. The capacitance was calculated from the discharge curve of the charge/discharge curves, using Equation \ref{eq:Cap1}.

To measure how well the slurry coating improved charge retention of the electrodes, cells were charged up to 1.5 V using a constant current of 0.1 A g\textsuperscript{-1}. Instead of being discharged at the same current upon reaching 1.5 V, the cell was left in its charged state with no applied current, while still measuring voltage continuously for 24 hours. This allowed the self-discharge mechanisms to occur and for observation on how this affected cell voltage. 

Brunauer-Emmett-Teller (BET) gas adsorption has emerged as the most commonly used technique to measure specific surface area (SSA) among supercapacitor electrode materials \cite{chmiola_anomalous_2006, zhu_carbon-based_2011, brunauer_adsorption_1938, sing_use_2001}. This was achieved using a Micromeritrics Flowsorb II 2300 gas adsorption machine with voltage reading through a Megger M8057 multimeter. Samples were degassed at 115 \textsuperscript{o}C for over one hour before measurements were taken to ensure minimal moisture content. Adsorption readings were taken at 77 K by immersing the sample tube in a liquid nitrogen dewar, while desorption readings were recorded by warming the sample back up to room temperature in water. 

The molecular sieves used as the zeolite source in this work have a typical aluminosilicate framework containing sodium, aluminium, silicon, and oxygen. As such, energy-dispersive X-ray spectroscopy (EDXS) can be used in tandem with scanning electron microscopy (SEM) to determine the composition and presence of elements in the sample, along with it's morphology. All microscopy measurements were performed on a JEOL 6600LV electron microscope.

\section{Results and discussion}
\label{S:3}
\begin{figure}[hbt]
\centering\includegraphics[width=0.6\linewidth]{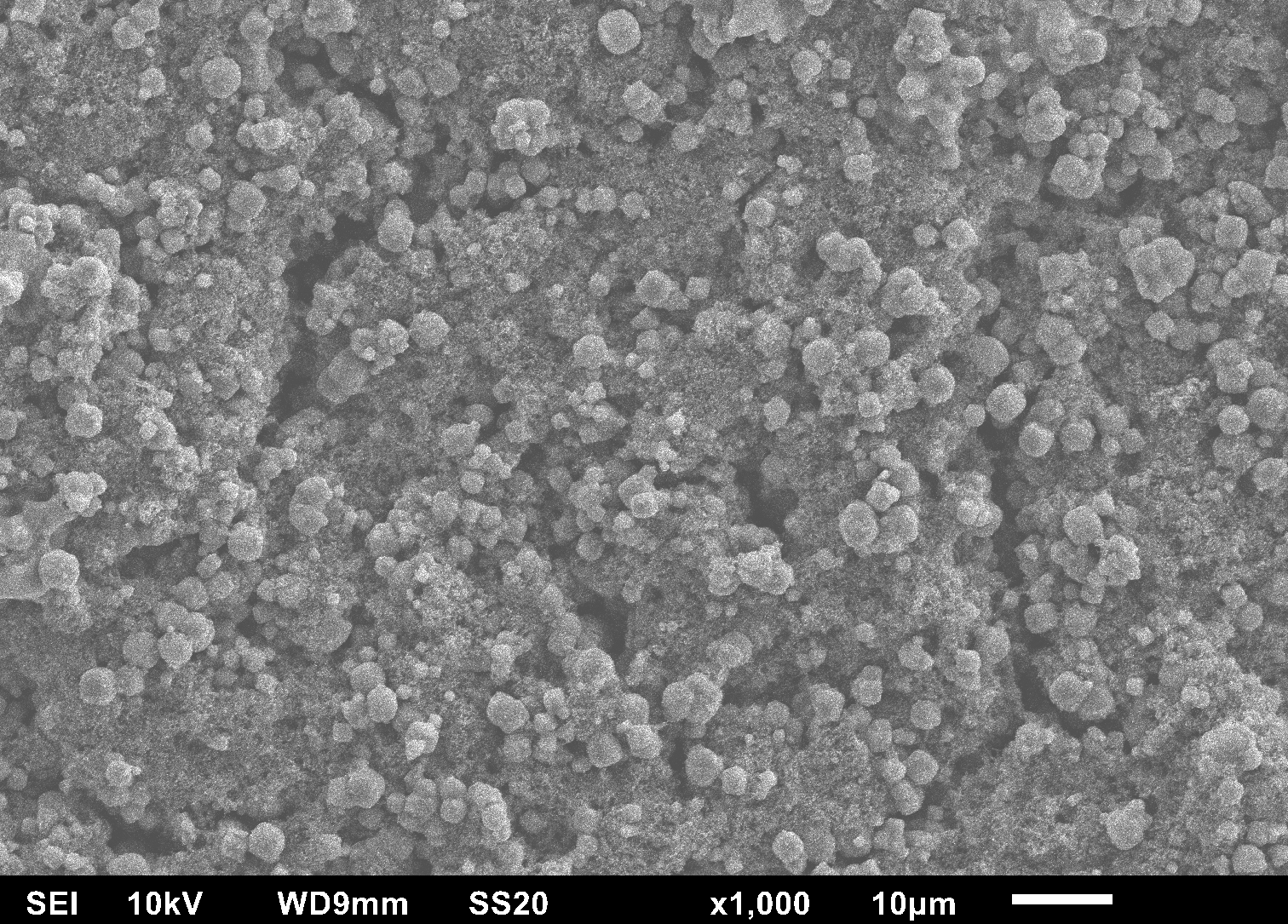}
\caption{SEM micrograph of a 70 wt\% zeolite-slurry coated carbon electrode at 500x magnification.}
\label{SEM-image}
\end{figure}

The 70 wt\% zeolite slurry can be seen in Figure \ref{SEM-image} to have a homogenous dispersion of particles throughout the slurry. These particles were determined to be zeolite particles via EDXS, as shown in Figure 3, where the elemental constituents of oxygen, sodium, aluminium, and silicon were found in the same locations. These locations overlap with the lighter coloured cube-like particles in the backscattered image. Furthermore, the carbon EDXS imaging shows lowered amounts of carbon in areas where these particles appear, confirming the presence of well dispersed zeolite particles in a carbon slurry.

\begin{figure}[hbt]
\centering\includegraphics[width=0.85\linewidth]{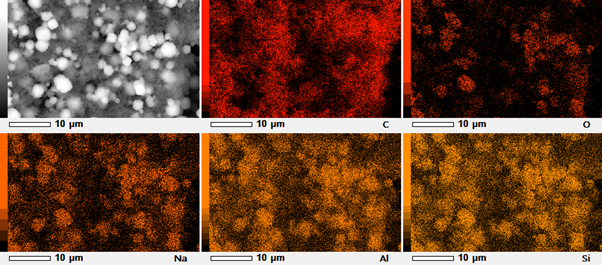}
\caption{Backscattered image (top-left) and EDXS elemental distribution data showing the presence and position of different elements (rest) for the zeolite-slurry coated carbon electrode.}
\label{EDS-image}
\end{figure}

The 4 Å molecular sieves were found to have an average specific surface area of 3.28 m\textsuperscript{2} g\textsuperscript{-1} as powder, and then when mixed into the 70\% slurry and coated on carbon paper, this increased to 5.43 m\textsuperscript{2} g\textsuperscript{-1}. This increase is likely due to the carbon black additive, as the carbon paper did not provide enough specific surface area to be detected by the instrument, but Sakai \textit{et al.} \cite{sakai_carbon_2018} calculated this to be 0.3 m\textsuperscript{2} g\textsuperscript{-1}. Super P carbon black has been found to have a specific surface area of $\sim$60 m\textsuperscript{2} g\textsuperscript{-1} \cite{spahr_development_2011}. These values are for the surface accessible by N\textsubscript{2} molecules (molecular diameter of 3.62 Å) \cite{eliad_ion_2001} and so it was assumed that the K\textsuperscript{+} and Cl\textsuperscript{-} ions (hydrated ionic diameter of 3.31 and 3.32 Å respectively) \cite{zhong_review_2015} will be able to access a similar surface area.

Since the carbon paper had very little surface area and contributed relatively negligible current density (see supporting information), and therefore capacitance, compared to the overall slurry-coated electrode, it was assumed that the slurry provided all of the electrochemical properties of the measured electrodes. This meant that the applied constant current, measured in A g\textsuperscript{-1}, could be calculated using the mass of the slurry coating for subsequent CDC and charge retention measurements.

\begin{figure}[hbt]
\centering
\includegraphics[width=0.65\linewidth]{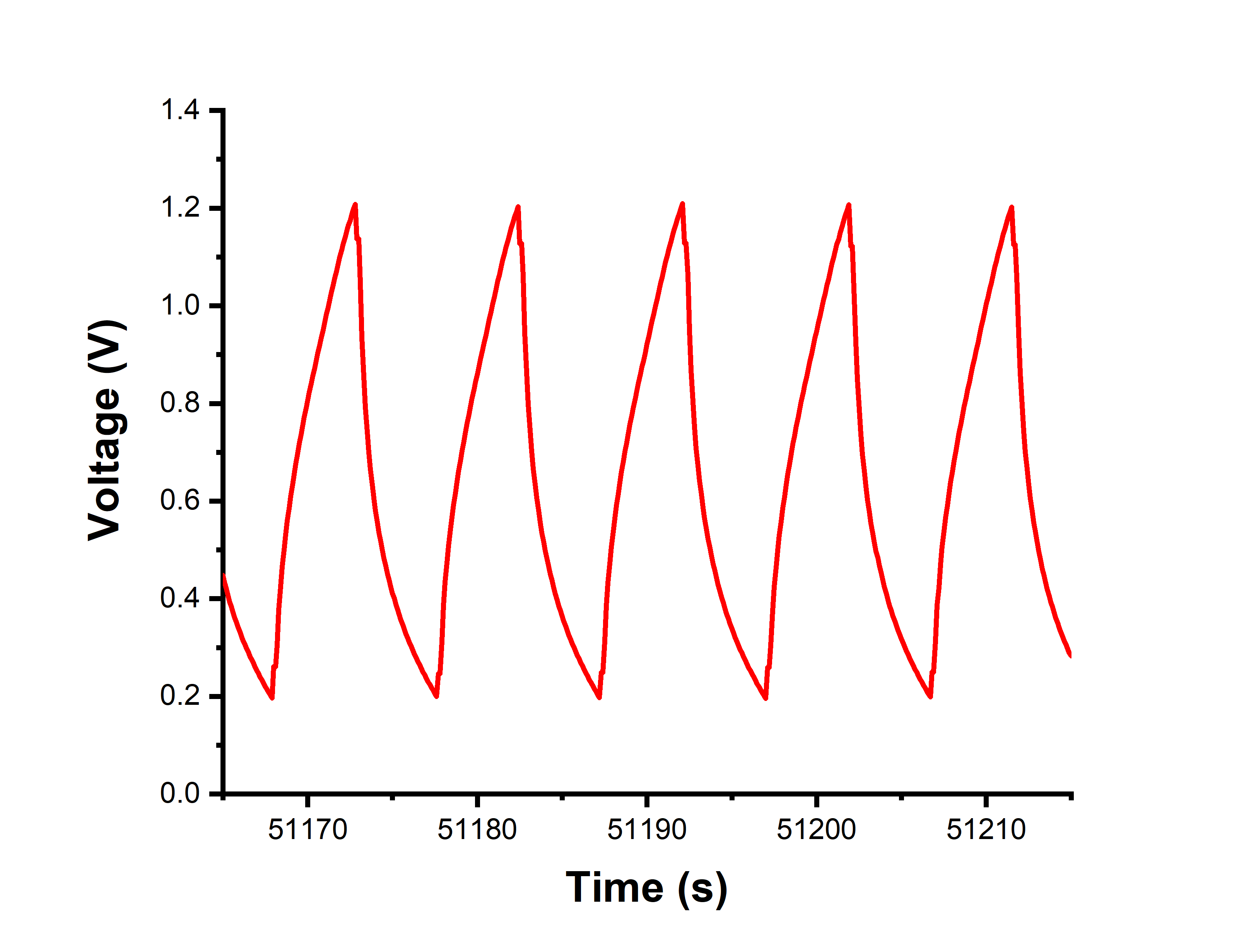}
\caption{Galvanostatic charge/discharge curves for 70\% zeolite slurry-coated supercapacitor cell, taken from midway through a 10000-cycle procedure, using constant current of 0.5 A g\textsuperscript{-1} (0.83 mA).}
\label{CDC-ZSC}
\end{figure}

CDC measurements for the supercapacitor cells with zeolite slurry-coated electrodes showed favourable performance. As demonstrated in Figure \ref{CDC-ZSC}, a symmetric, triangular shape in the charge/discharge curves is characteristic of a double-layer formation and capacitive properties while the consistent spacing between peaks is indicative of high stability. Specific capacitance was calculated as the average value from a 1000-cycle run using the discharge portion of the curve with Equations \ref{eq:Cap1} and \ref{eq:Cap2} and was found to be 17.25 F g\textsuperscript{-1}.

\begin{figure}[hbt]
\centering\includegraphics[width=0.65\linewidth]{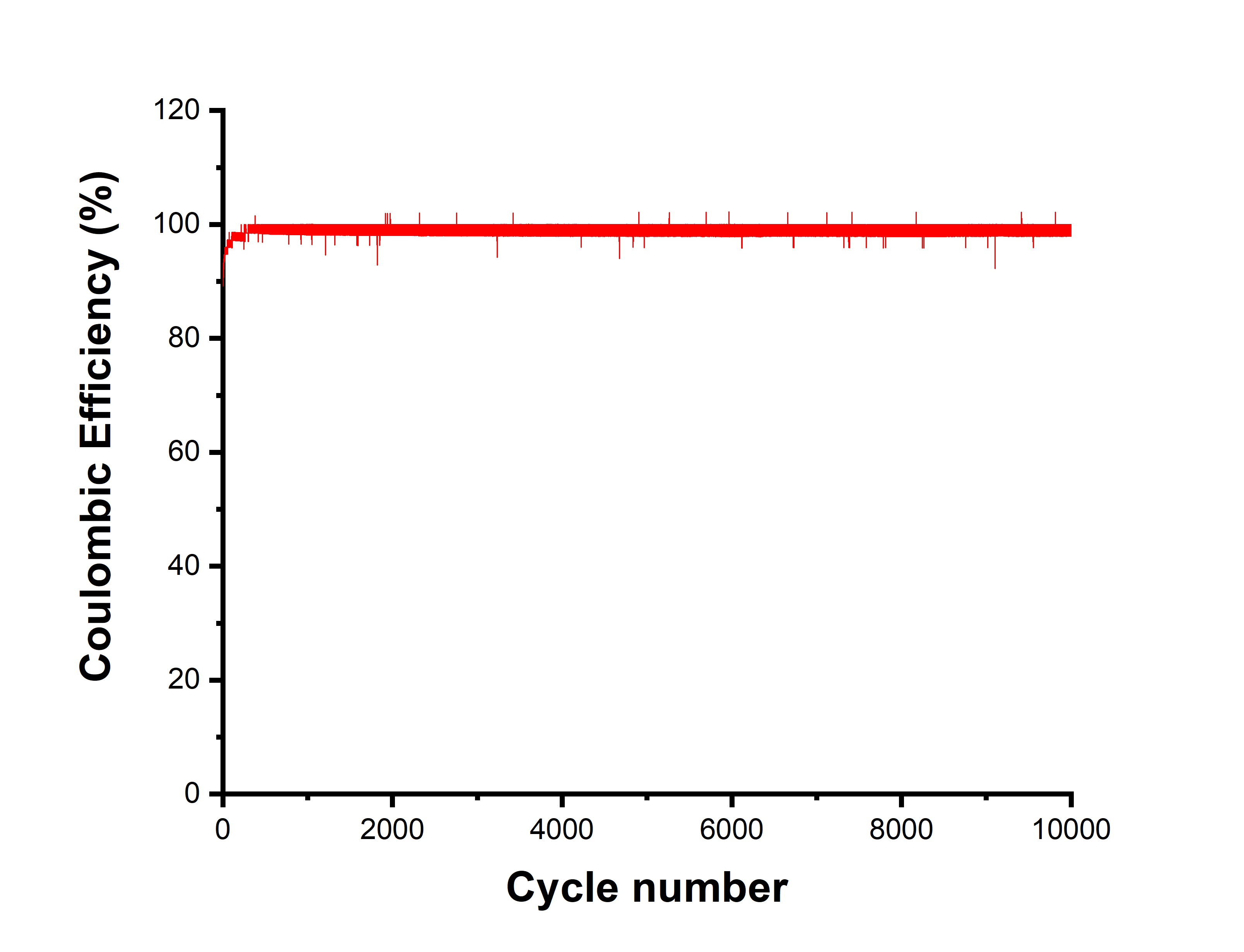}
\caption{Coulombic efficiency for 10,000 charge-discharge cycles at a constant current of 0.5 A g\textsuperscript{-1} (0.83 mA). Note this procedure was performed after over 15,000 previous charge/discharge cycles.}
\label{CE-ZSC}
\end{figure}

Several thousand charge/discharge cycles were performed, including a 5000-cycle run and a 10,000-cycle run, and it was found that the capacitance dropped over the course of the cycling procedures. After over 18,000 charge-discharge cycles at 0.5 A g\textsuperscript{-1}, the capacitance was calculated as 10.16 F g\textsuperscript{-1}. This meant that if the supercapacitor cell was charged and discharged once per day for half a century, almost 60\% of its capacitance would be retained. This is reflected in the high coulombic efficiency on the device, shown in Figure \ref{CE-ZSC}.

\begin{figure}[hbt]
\centering\includegraphics[width=0.65\linewidth]{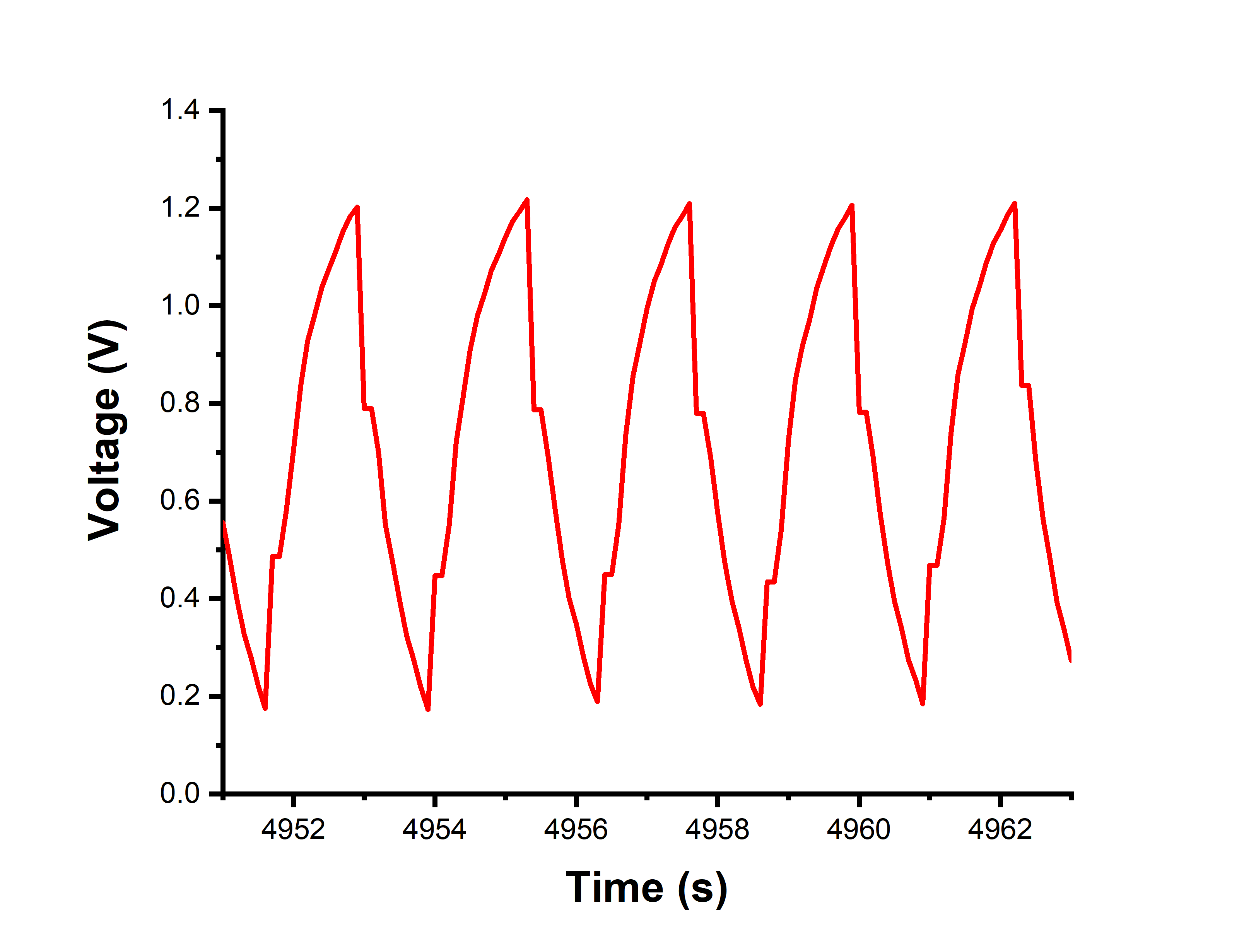}
\caption{Galvanostatic charge/discharge curves for uncoated carbon paper supercapacitor cell, taken from midway through a 5000-cycle procedure, using constant current of 0.1 A g\textsuperscript{-1} (0.56 mA).}
\label{CDC-SC}
\end{figure}

In comparison, the uncoated electrodes built up and released charge at a relatively high rate, likely due to the low surface area, which meant that its full capacity was reached quickly. A lower constant current of 0.1 A g\textsuperscript{-1} was required to yield a comparable charge/discharge curve, as using a constant current of 0.5 A g\textsuperscript{-1} gave a charge/discharge cycle time of about 0.7 s; too fast for accurate instrument reading or realistic application. In Figure \ref{CDC-SC}, it can be seen that 0.1 A g\textsuperscript{-1} gave a charge-discharge cycle time of about 2 s.

Uncontrolled charging and discharging was observed for the uncoated supercapacitor, seen in Figure \ref{CDC-SC}, where the voltage spikes above and below 1.2 and 0.2 V respectively. Along with the uncontrolled charge/discharge voltages, significant IR drops were observed in every CDC for uncoated electrodes. This was a drop in over one third of the max voltage (from over 1.2 to under 0.8 V), indicating a large internal cell resistance, especially as compared with the slurry-coated cell’s IR drop of about 5\% (from 1.2 to above 1.1 V).

Capacitance for the uncoated electrode was determined as the average value from discharge curves of a 1000-cycle CDC procedure and was found to be 0.40 F g\textsuperscript{-1}. This shows that the addition of surface area and a porous surface increased the supercapacitive properties of the cell. Improvements were similarly found in the slurry-coated electrodes for charge retention, shown in Figure 7.

\begin{figure}[hbt]
\centering\includegraphics[width=0.65\linewidth]{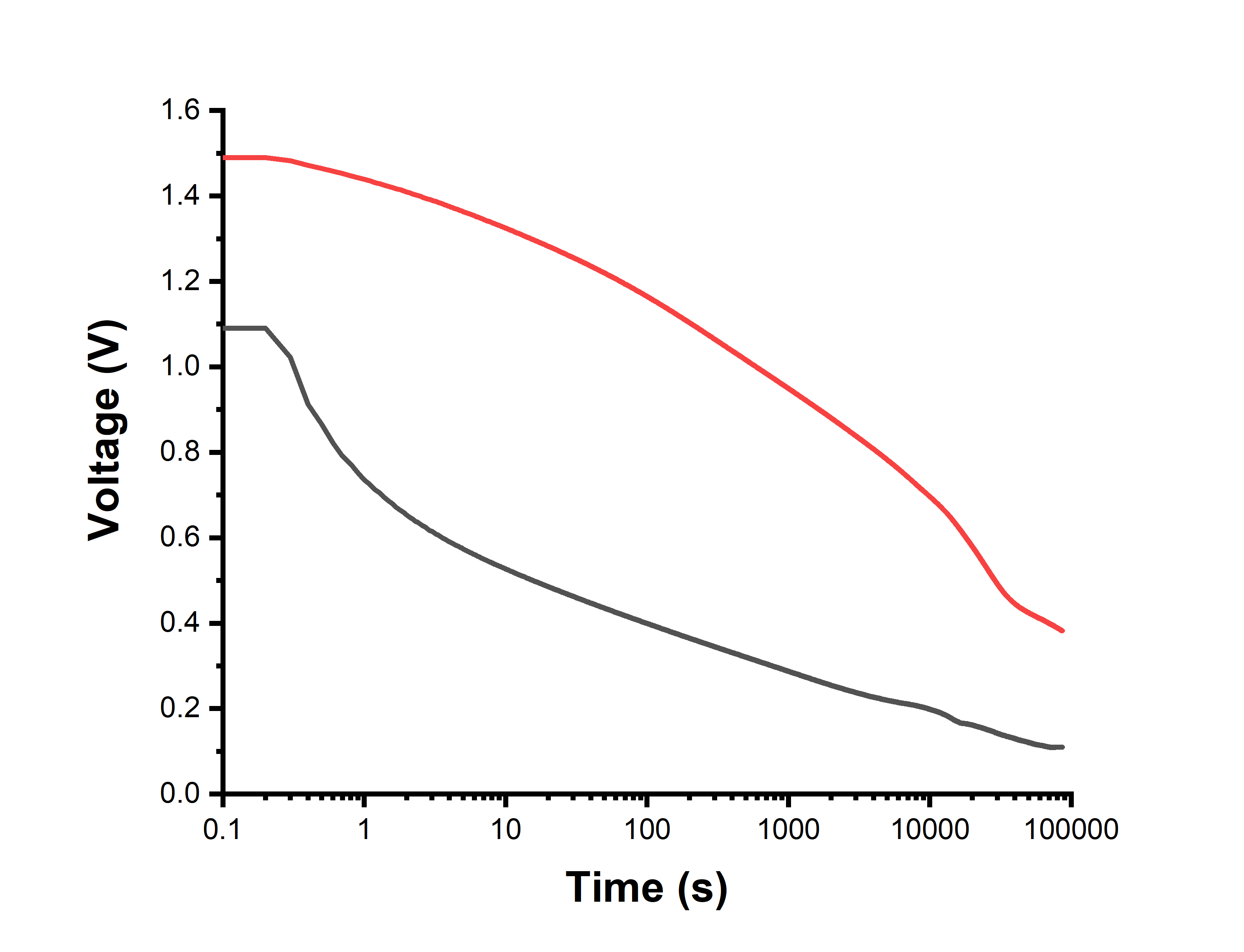}
\caption{The measured voltage drop for the 70\% zeolite slurry-coated supercapacitor cell (red) and the uncoated carbon paper electrode (black), charged up to 1.5 V at a constant current of 0.1 A g\textsuperscript{-1}, with time displayed as a logarithmic scale.}
\label{CR-comparison}
\end{figure}

Overall, a significant improvement can be seen in the cell's charge retention over the course of 24 hours, with the zeolite slurry-coated supercapacitor cell voltage still not plateauing by the end of the measured period. In contrast, the uncoated carbon paper cell began to plateau after about 30,000 seconds (just over 8 hours) at a voltage of 0.11 V. While this is not the open cell voltage of 0 V, it would require an extended amount of time for absolutely all charge to dissipate from the electrodes, and so the plateau can be interpreted as essentially uncharged.

The zeolite slurry-coated supercapacitor cell maintained a voltage of 0.38 V after 24 h, while the uncoated carbon paper supercapacitor had a voltage of 0.11 V. For an initial cell voltage of 1.5 V, this is a difference of 25.48\% charge retention over 24 h compared to 7.31\%, which is an enhanced retention of 349\%. The slurry consisted of conductive carbon black and porous molecular sieves. We proposed that the higher conductivity and capacitance of the electrode is due to the carbon black additive, whereas the charge retention capabilities can be attributed to the zeolite molecular sieves. Both slurry components can be seen to contribute to the observed electrode performance, as hypothesised.

Since we have an approximate linear relation in the CDC between voltage and current, the voltage is therefore proportional to the state of charge, and subsequently the number of adsorbed electrolyte ions. The self-discharge due to ion desorption can be written as a chemical reaction of A $\rightarrow$ B with A as the absorbed ion and B as the free ion (Refer to Figure \ref{fig_scheme}). This first-order reaction will have a rate constant ($k$) and some Gibbs free energy of activation ($\Delta G^\ddagger$), which can be derived from the rate constant based on transition state theory, using the Eyring equation:

\begin{equation}
k = \frac{\kappa k_B T}{h} e^{-\frac{\Delta G^\ddagger}{RT}}
\label{eq:Eyring}
\end{equation}

Where $\kappa$ is the transmission coefficient (often taken to be unity), $k_B$ the Boltzmann constant, $h$ the Planck constant, and $R$ the universal gas constant.
There are three main mechanisms of self-discharge, as identified and discussed in detail by Conway \textit{et al.}: Overpotential, diffusion controlled, and ‘leaky capacitor’. \cite{conway_diagnostic_1997} For the three identified mechanisms of self-discharge, a linear dependence was found due to overpotential through voltage on a logarithmic timescale. For a diffusion-controlled pathway of self-discharge the linear dependence is voltage on the square root of time, and for a leaky capacitor it is logarithmic voltage on time.

\begin{figure}[hbt]
\centering\includegraphics[width=0.65\linewidth]{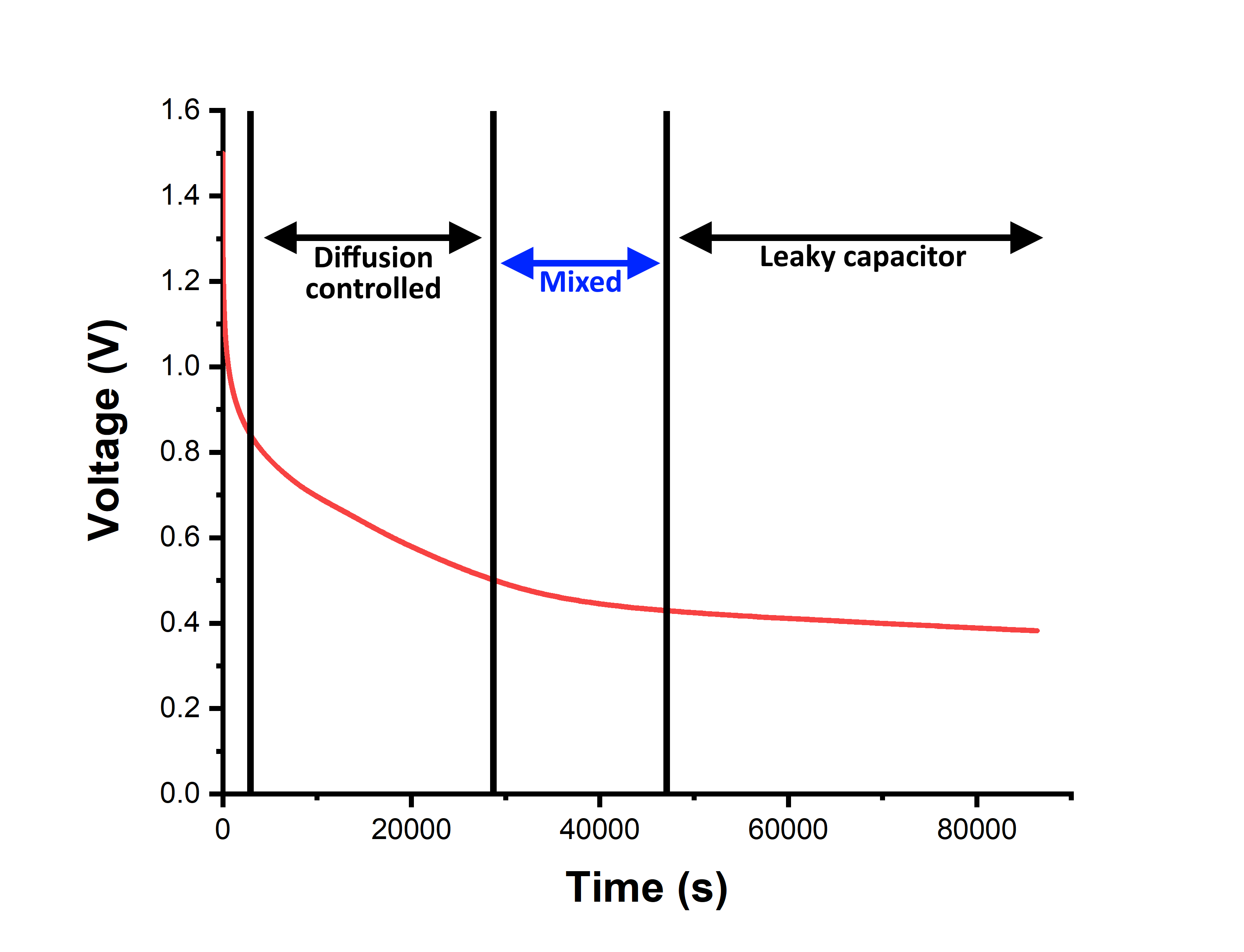}
\caption{Voltage vs$.$ time self-discharge curve with the dominant mechanisms of self-discharge for each time region labelled. A mix of the two mechanisms is observed between each one.}
\label{CR-annotated}
\end{figure}

These mechanisms often occur simultaneously but Ricketts and Ton-That found that there can be regions of time where one mechanism is seen to dominate \cite{ricketts_self-discharge_2000}, and indeed for our case in Figure \ref{CR-annotated}, the self-discharge observed was diffusion controlled early on, but then transitioned to leaky capacitor behaviour until the voltage plateau. The third mechanism, that of a leaky capacitor, has the same linear relationship as for a first-order reaction, and so will have a gradient equal to $-k$. This value can be substituted into Equation \ref{eq:Eyring} and used to find a value for $\Delta G^\ddagger$.

\begin{figure}[hbt]
\centering\includegraphics[width=0.65\linewidth]{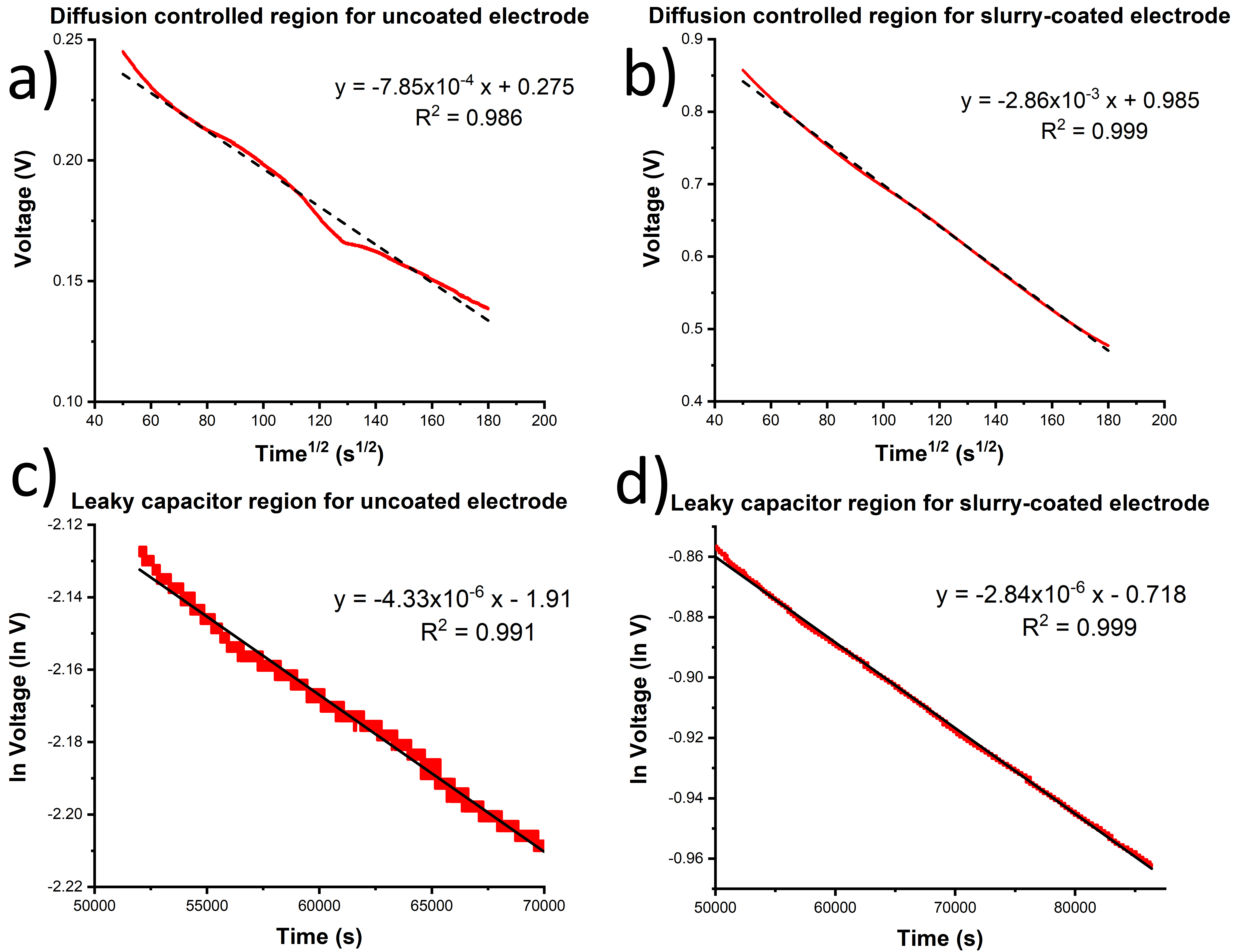}
\caption{Transformed data for the self-discharge of the uncoated (a, c) and coated (b, d) electrodes}
\label{CR-mechanisms}
\end{figure}

The transformed plots are shown in Figure \ref{CR-mechanisms} for the relevant time period. After about 40 minutes, the ion desorption entered a regime dominated by diffusion controlled self-discharge (Figure \ref{CR-mechanisms}a, \ref{CR-mechanisms}b), which lasted for roughly 8 hours until the leaky capacitor mechanism began to become significant. This mix of processes occurred for about 5 hours, until the voltage had a solely exponential decay rate, shown in Figure \ref{CR-mechanisms}c and \ref{CR-mechanisms}d. The voltage plateaued after this exponential decay, noted at 19.5 hours for the uncoated electrode, but not encountered for the coated electrode.

It can be seen that the slope of the electrodes without the zeolite slurry is steeper in both cases. This quantitatively showed that desorption occurred slower for the zeolite slurry electrode, but using Equation \ref{eq:Eyring} qualitatively showed this. At a temperature of 293.15 K, the rate constant for the uncoated electrode of $k = 4.33 \times 10^{-6}$ s$^{-1}$ yielded a Gibbs free energy of activation of $\Delta G^\ddagger = 101,840$ J mol$^{-1}$, while for the zeolite coated electrode, $k = 2.84 \times 10^{-6}$ s$^{-1}$ gave $\Delta G^\ddagger = 102,868$ J mol$^{-1}$. This is an increase in activation energy of over 1 kJ mol$^{-1}$, simply through the addition of the zeolite slurry.

\section{Conclusion}
\label{S:4}
A conceptually new supercapacitor was constructed in air, on a benchtop, using low-cost, Earth-abundant materials with aqueous potassium chloride electrolyte that exhibited 100\% coulombic efficiency and excellent stability over the equivalent of almost 50 years of testing and cycling. The supercapacitor was designed from first principles, aimed at increasing the free energy barrier that charge carriers have to overcome in order to desorb from the electrodes. Given the low surface area of the carbon source and the insulating nature of the zeolites, the stated values could easily be improved upon. In this proof-of-concept study, it has been shown that this design can provide competitive stability and capacitance, while also increasing the charge retention capabilities.

Our focus was on charge retention without specific regard to capacitance, but despite this, a capacitance value of 17.25 F g\textsuperscript{-1} was achieved for a carbon-based zeolite-slurry electrode. This is comparable with carbonaceous slurries and noteworthy considering that the slurry consisted of 70\% molecular sieves, an insulator rather than a conductor. Improved charge retention was also observed, with more than 25\% of the initial charge being retained over the measured 24 hours.

In conclusion, we have designed a new type of supercapacitor, based on the hypothesis that free desorption energies determine the charge retention in supercapacitors. In a proof-of-concept study, we constructed a model supercapacitor from first principles and were able to verify our hypothesis. Despite not aiming for high performance, our supercapacitors showed comparable performance metrics as other carbon-based systems.

\bibliographystyle{unsrt}  
\bibliography{ZoteroBBT_supercaps}  


%
%
%

\end{document}